%% file: LF.tex
\documentclass[traditabstract]{aa}

\usepackage{epsfig,caption}
\usepackage{multirow}
\usepackage{graphics}
\usepackage[latin1]{inputenc}
\usepackage{natbib}

\begin{document}

\title{\bf Luminosity Function from dedicated SDSS-III and MMT data of quasars in $0.7<z<4.0$ selected with a new approach
}
\author{N. Palanque-Delabrouille\inst{1} \and Ch. Magneville \inst{1} \and Ch. Y\`eche\inst{1} \and S. Eftekharzadeh\inst{2} \and   A. D. Myers
\inst{2,3} \and  P. Petitjean\inst{4} \and I.~P\^aris\inst{4,5} \and  E. Aubourg\inst{6} \and I. McGreer\inst{7} \and X. Fan\inst{7} \and A. Dey\inst{8} \and D. Schlegel\inst{9} \and S. Bailey\inst{9} \and  D. Bizayev\inst{10} \and A. Bolton\inst{11} \and K. Dawson\inst{11} \and G. Ebelke\inst{10} \and J. Ge\inst{12}
\and E. Malanushenko\inst{10} \and V. Malanushenko\inst{10}
 \and D. Oravetz\inst{10} \and K. Pan\inst{10} \and N. P. Ross\inst{9}  \and D.P.  Schneider\inst{13} \and E. Sheldon\inst{14} \and A. Simmons\inst{10} 
\and J. Tinker\inst{15} \and M. White\inst{9} \and Ch. Willmer\inst{7}
} 

\institute{CEA, Centre de Saclay, Irfu/SPP,  F-91191 Gif-sur-Yvette, France
\and Department of Physics and Astronomy, University of Wyoming, Laramie, WY 82071, USA
\and Max-Planck-Institut f\"{u}rAstronomie, K\"{o}nigstuhl 17, D-69117 Heidelberg, Germany 
\and Universit\'e Paris 6, Institut d'Astrophysique de Paris, CNRS UMR7095, 98bis blvd Arago, F-75014 Paris, France 
\and Departamento de Astronom\'ia, Universidad de Chile, Casilla 36-D, Santiago, Chile 
\and APC, 10 rue Alice Domon et L\'eonie Duquet, F-75205 Paris Cedex 13, France 
\and Steward Observatory, University of Arizona, Tucson, AZ 85721, USA 
\and  National Optical Astronomy Observatory, Tucson, AZ85726-6732,USA 
\and Lawrence Berkeley National Laboratory, 1 Cyclotron Road, Berkeley, CA94720, USA
\and Apache Point Observatory, P.O. Box 59, Sunspot, NM 88349-0059, USA
\and University of Utah, Dept. of Physics \& Astronomy, 115 S 1400 E, Salt Lake City, UT 84112, USA 
\and Department of Astronomy, University of Florida, 211 Bryant Space Science Center, P.O. Box 112055, Gainesville, FL 32611, USA
\and Department of Astronomy and Astrophysics, The Pennsylvania State University, University Park, PA 16802\\
Institute for Gravitation and the Cosmos, The Pennsylvania State University, University Park, PA 16802
\and Brookhaven National Laboratory, Bldg 510, Upton, NY  11973, USA 
\and Center for Cosmology and Particle Physics, New York University, New York, NY 10003, USA 
}
\date{Received xx; accepted xx}
\authorrunning{N. Palanque-Delabrouille et al.}
\titlerunning{Luminosity Function from dedicated SDSS-III and MMT data }
\abstract{We present a measurement of the quasar luminosity function in the range $0.68<z<4$ down to extinction corrected magnitude $g_{\rm dered}=22.5$, using a simple and well understood target selection technique based on the time-variability of quasars. The completeness of our sample was derived directly from a control sample of quasars, without requiring complex simulations of quasar light-curves or colors. A total of 1877 quasar spectra were obtained from dedicated programs on the Sloan telescope (as part of the SDSS-III/BOSS survey) and on the Multiple Mirror Telescope. They allowed us to derive the quasar luminosity function. It agrees well with  results previously published in the  redshift range $0.68<z<2.6$.  Our deeper data  allow us to extend the measurement to $z=4$. We measured quasar densities to $g_{\rm dered}<22.5$, obtaining  30~QSO per ${\rm deg}^{2}$ at $z<1$, 99~QSO per deg$^{2}$ for $1<z<2.15$, and 47~QSO per deg$^{2}$ at $z>2.15$. Using pure luminosity evolution models, we fitted our LF  
measurements and  predicted quasar number counts as a function of redshift and observed magnitude.
 These predictions are useful inputs for future cosmology surveys such as those relying on the observation of quasars to measure baryon acoustic oscillations. }
\maketitle

\section{Introduction}

The measurement of the Baryon Acoustic Oscillation (BAO) scale~\citep{bib:eisenstein05, bib:eisenstein07} relies on large samples of objects selected with an unbiased method. To probe the distant Universe, quasars appear to be one of  the sources of choice, since they are both among the brightest extragalactic objects, and expected to be present at sufficiently high density.  

The selection of quasars to redshift $z\sim 4$ and magnitude $g\sim23$, which is the objective of current and future cosmology projects dedicated to BAO studies in the distant Universe, is a major challenge. Traditional selections relying on quasar colors for several broad optical bands \citep{bib:schmidt83, bib:croom01, bib:richards02, bib:richards04, bib:richards09, bib:croom09, bib:Bovy11, bib:Bovy12} present serious drawbacks for the selection of quasars at redshifts near z$\sim$2.7, which occupy similar regions of $ugriz$ color-space as the more numerous white dwarfs and blue halo stars~\citep{bib:Fan, bib:richards02, bib:worseck}.  To circumvent this difficulty, \cite{bib:palanque} developed a selection algorithm relying on the time variability of quasar fluxes. This technique was tested in 2010 as part of the BOSS survey~\citep{bib:ross12}. It was shown to increase by 20 to 30\% the density of identified quasars, and, in particular, to effectively recover additional quasars in the redshift range $2.5<z<3.5$.

Here we use this variability-based selection to measure the  quasar luminosity function to extinction-corrected magnitude $g_{\rm dered} = 22.5$ and redshift $z= 4$ from two sets of dedicated observations. For the first set of  data, we took advantage of the re-observation, as part of the SDSS-III/BOSS survey~\citep{bib:eisenstein11,bib:dawson12}, of a $14.5\,{\rm deg}^2$ region in Stripe 82 (the SDSS Southern equatorial stripe). The second set of data used the Hectospec multi-object spectrograph~\citep{bib:fabricant05} on the Multiple Mirror Telescope (MMT) and covered $4.7\, {\rm deg}^2$.  The quasar samples being selected with only minimal color constrains, they are expected to be highly complete and do not suffer from the usual biases in their redshift distribution induced by color selections. The selection algorithms of both samples are well understood, and can be applied to large control samples of already identified quasars in order to compute the completeness of the method. With this strategy, all corrections can be derived from the data themselves and do not require any modeling of quasar light curves or colors.

The outline of the paper is as follows. In Sec.~\ref{sec:target} we explain the strategy for the selection of the targets and its specific application to the BOSS and the MMT observations.  In Sec.~\ref{sec:eff}, we describe the different contributions to the global completeness correction for both sets of data and present the quasar control sample used to derive them. The raw and the completeness-corrected quasar number counts are given in Sec.~\ref{sec:results}, where we also present several cross-checks of the results obtained. The quasar luminosity function in $g$ derived from these data is given in Sec.~\ref{sec:QLF}.

\section{Target selection} \label{sec:target}

While the basis of the target selection algorithm is the same for the two components of our program, it was applied with different thresholds to obtain the targets for the BOSS and the MMT observations.
The BOSS component was indeed designed to identify a large number of quasars to a magnitude limit $g\sim 22.5$ corresponding to the limit of BOSS spectroscopy at typical exposure times. The MMT data were primarily designed to complete the sample to fainter magnitudes ($g\sim 23$), since the telescope has a 6.5~m primary mirror compared to the 2.5~m primary of the Sloan telescope. In addition, the BOSS sample was restricted to point sources, while the MMT sample was also used to recover quasars lying in extended sources. 

\subsection{Target selection algorithm} \label{sec:TSalgorithms}

As a detailed description of the variability selection can be found in~\cite{bib:palanque}, we only summarize here the major steps of our algorithm.  

For each source, lightcurves were computed from the data collected by SDSS using  the dedicated Sloan Foundation 2.5~m telescope~\citep{bib:gunn06}. A mosaic CCD camera~\citep{bib:gunn98} imaged the sky in the $ugriz$ bandpasses~\citep{bib:fukugita96}. The imaging data were processed through a series of pipelines~\citep{bib:stoughton02} which performed astrometric calibration, photometric reduction and photometric calibration. 

The starting source list was built from images that resulted from the co-addition~\citep{bib:annis12} of SDSS single-epoch images.\footnote{We used the Catalog Archive Server (CAS) interface (http://casjobs.sdss.org) to recover both the Stripe 82 coadd and the single epoch information.} The completeness of the coadd catalog reaches 50\% at a magnitude $g=24.6$ for stars and $g=23.5$ for galaxies. The $g$ magnitude mentioned throughout  the paper results from a PSF-fit on the coadd images, and effectively represents the mean magnitude of a variable source. The source morphology (point-like or extended) is also determined from these coadds.

\begin{figure}[h]
\begin{center}
\epsfig{figure=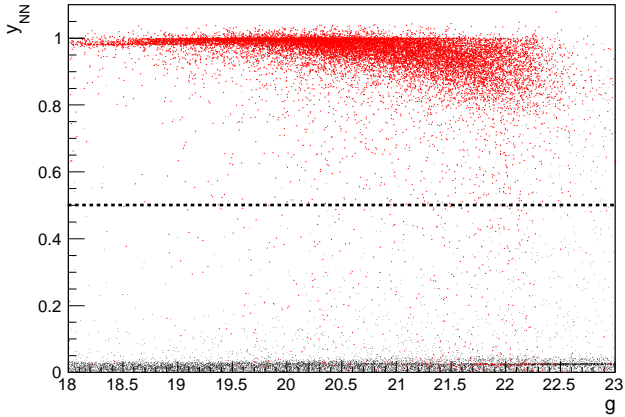,width = \columnwidth} 
\caption[]{Output of the variability Neural Network as a function of $g$ magnitude for a sample of known stars (small black dots near $y_{NN}= 0$) and for known quasars (larger red dots at $y_{NN}\sim1$).  BOSS targets and the MMT point-source targets are required to pass the criterion $y_{NN}>0.5$.}
\label{fig:NNvar_g}
\end{center}
\end{figure}

The lightcurves of our sources contained on average 52 individual epochs spread over 7 years. They were used to compute two sets of parameters that characterize the source variability.  These parameters are:
\begin{itemize}
\item { the five $\chi^2$s for the fit of the lightcurve in each of the $ugriz$ filters by a constant ($\overline{m}$),
\begin{equation}
\chi^2 = \sum_i \left( \frac { m_i - \overline{m} } {\sigma_i} \right) ^2\, , 
\label{eq:var_chi}
\end{equation}
where the sum runs over all observations $i$.}
\item two parameters, an amplitude $A$ and a power $\gamma$ as introduced by \cite{bib:schmidt10}, that characterized the variability structure function $\mathcal V(\Delta t_{\rm ij})$, i.e.  the change in magnitude $\Delta m_{\rm ij}$ as a function of time lag $\Delta t_{\rm ij}$ for any pair $ij$ of observations, with 
\begin{eqnarray}
\mathcal V(\Delta t_{\rm ij}) &=& |\Delta m_{\rm i,j}| - \sqrt{\sigma_{\rm i}^2 + \sigma_{\rm j}^2} \\
& = & A\times (\Delta t_{\rm ij})^\gamma\,.
\label{eq:var_Agamma}
\end{eqnarray}
\end{itemize}
For each source, a neural network then combined  the five $\chi^2$, the power $\gamma$ (common to all filters) and the amplitudes $A_g$, $A_r$ and $A_i$ for the three filters least affected by noise and observational limitations ($gri$), to produce an estimate of quasar-like variability.  An output $y_{\rm NN}$ of the neural network near $0$ designated non-varying objects, as is the case for the vast majority of stars, while an output near $1$ indicated a quasar (cf. Fig.~\ref{fig:NNvar_g}).  

This technique has been applied by BOSS for the selection of $z>2.2$ quasars in Stripe 82, down to $g\sim 22$~\citep{bib:palanque}. As was clearly illustrated by this study, this approach presents the advantage of being highly complete (the selection reached the unprecedented quasar completeness of 90\%), even for a sample purity of 92\%, higher than for typical selections based on quasar colors. In addition, this variability-based selection was shown to overcome the drawbacks of color-based methods and to recover quasars near redshift $z\sim3$ that are systematically missed with traditional selections~\citep{bib:richards02}.

Here, we applied this technique to fainter magnitudes, with the aim of detecting quasars to $g\sim23$. The magnitude dependence of the output of the variability neural network is illustrated in Fig.~\ref{fig:NNvar_g}. Requiring an output $y_{NN}> 0.5$ selects 95\% of the sample of known quasars  with $18<g< 23$ (cf. the description of this control sample in section~\ref{sec:knownsample}). Even when restricting to faint quasars with a magnitude $g>22$, $88\%$ of them still pass this threshold.

\begin{figure}[h]
\begin{center}
\epsfig{figure=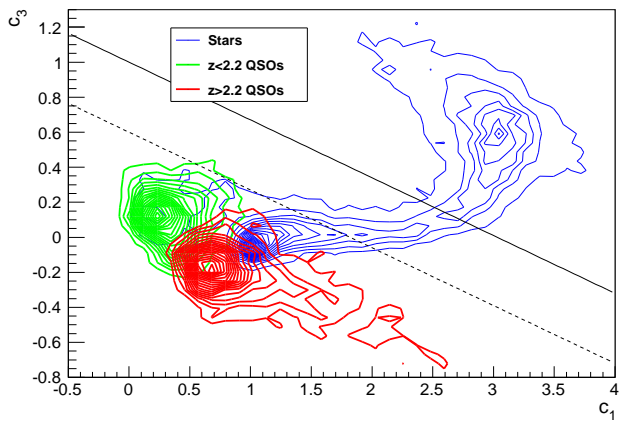,width = \columnwidth} 
\caption[]{ Locus of stars (upper blue contours), $z<2.2$ quasars (lower left green contours) and $z>2.2$ quasars (lower right red contours) in the $c_3$ vs. $c_1$ color-color plane. The upper solid line corresponds to the color cut $c_3<1.0 - c_1 / 3$ (loose, for point sources) and the lower dashed line to $c_3<0.6- c_1 / 3$ (strict, for extended sources). }
\label{fig:c1c3}
\end{center}
\end{figure}

In addition to the variability-based selection, a loose color constraint was used to reject a region of color-space mostly populated by stars, in order to reduce the fraction of stellar contaminants in the target list. The cut consisted in requiring $c_3<a-  c_1 / 3$, where  $c_1$ and $c_3$ are defined as in~\cite{bib:Fan} by 
\begin{eqnarray}
c_1 &=& 0.95(u-g)+0.31(g-r)+0.11(r-i) \, ,\nonumber \\
c_3 &=& -0.39(u-g)+0.79(g-r)+0.47(r-i)\, .
\label{eq:c1c3} \end{eqnarray} 
The parameter $a$ can take either of two values: $a=1.0$ for a very loose color cut, or $a=0.6$ for a tighter color constraint  (cf. Fig.~\ref{fig:c1c3}).  About 100\% of $z<2.2$ and 98\% of $z>2.2$ known quasars (resp. 95\% and 93\%) pass the condition with $a=1.0$ (resp. $a=0.6$).

\subsection{The BOSS sample}\label{sec:TS_BOSS}

The BOSS fields dedicated to this high density quasar survey were part of Stripe 82 and located at $317^\circ < \alpha_{\rm J2000} < 330^\circ $ and $0^\circ <\delta_{J2000}<1.25^\circ $. 

The starting source list consisted of all point sources in this area passing the usual BOSS quality cuts (on photometry, pixel saturation, source blending etc.) as described in the Appendix of~\citet{bib:ross12}, as well as the loose color cut (with $a=1.0$) mentioned in the previous section, yielding about 5200 objects per ${\rm deg}^2$. 

All objects passing $y_{NN}>0.5$ were selected as targets. The resulting list contained $\sim 270$ targets per ${\rm deg}^2$. To increase the identification rate of new targets, we removed from this list all targets that had already been observed by BOSS~\citep{bib:ahn12}. They consisted almost exclusively of spectroscopically identified quasars with redshifts between 0 and 5, totaling  a density of $\sim 30$ quasars per square degree. This reduced the list to $\sim 240$ objects per ${\rm deg}^2$. 

In case of fiber collisions during the tiling procedure, priorities were set on the targets according to their magnitude, giving higher priority to the brightest objects since these are more likely to obtain an accurate identification. An identical maximal priority was attributed to all targets with $g<22.7$ (not corrected for extinction as the quantity of interest here is the actual observed flux of the object), and lesser priorities for fainter magnitudes.

Seven partially overlapping BOSS half-plates were allocated to the project (numbered 5141 to 5147 from low to high $\alpha_{\rm J2000}$), covering a total of 14.5 ${\rm deg}^2$. The other plate halves were used for the standard BOSS survey and are not included in this analysis. Plate 5141 was exposed for over 2 hours (ie. twice the normal time), without significant increase in the number of quasars identified. The subsequent plates were thus exposed for 1 hour. All seven plates were observed during July and September 2011.

\subsection{The MMT sample}\label{sec:TS_MMT}

For the sake of completeness, the starting list for the MMT sample was built from all SDSS sources in the area of the previous BOSS sample, whether point-like or extended and whether or not they passed the standard BOSS quality cuts, and passing the loose color cut with $a=1.0$ described in Sec.~\ref{sec:TSalgorithms}. 
This initial list consisted of $\sim 11300\,{\rm deg}^{-2}$ objects, 60\% of which were resolved (ie. ``extended", like galaxies), and 40\% of which were unresolved (ie. ``point-like'', like stars). 
As for the BOSS sample, all objects that have already obtained spectra with SDSS-III/BOSS were removed from the target list.  These were again mostly quasars.
In addition, we rejected targets with $g<22$ that were simultaneously selected for the BOSS sample, since the redshift determination efficiency of BOSS is close to unity at least to that magnitude. 

Most quasars are expected to be significantly more luminous than their host galaxy, making them appear point-like. We thus expect less quasars in the ``extended" sub-sample. The fraction of random objects artificially drifting into the region of large  $y_{NN}$ values being independent of the source morphology, however, the ratio of the number of quasars to the total number of selected targets is therefore smaller for extended sources. To counteract the effect of a more liberal morphological selection, extended sources were therefore subject to tighter constraints on both color and variability.
They were also given a lesser priority than the MMT point-like targets. 

The highest priority (priority A) was given to point-like targets with observed magnitude $g<23.0$, $y_{NN}>0.5$ and $c_3<1.0-0.33c_1$, leading to $\sim 130\,{\rm deg}^{-2}$ targets. 

Priority B was given to extended objects with $g<23.0$, $y_{NN}>0.8$ and $c_3<0.6-0.33\times c_1$. This led to 140~${\rm deg}^{-2}$ additional targets. 

Priority C (respectively D) was given to objects with $23.0<g<23.2$ passing the same conditions as the priority A (resp. B) targets, leading to 40~${\rm deg}^{-2}$ (resp. 30~${\rm deg}^{-2}$) additional targets. 
 
A total of 6 non-overlapping circular fields of 1~${\rm deg}$ in diameter were observed with MMT/Hectospec in the last trimester of 2011, with exposure times of 150 minutes each. The fields were located within the coverage of the BOSS sample of Sec.~\ref{sec:TS_BOSS}, in the area farthest from the Galactic plane, which is least contaminated by Galactic stars.  The region $326.5^\circ<\alpha_{\rm J2000}<328^\circ$ was avoided because of a higher  Galactic extinction on average ($A_g\sim 0.4$) than for the other MMT fields ($A_g\sim 0.2$ for $322^\circ<\alpha_{\rm J2000}<326.5^\circ$ and $A_g\sim 0.3$ for $328^\circ<\alpha_{\rm J2000}<330^\circ$), where the extinctions come from the maps of~\citet{bib:schlegel98}.

Because of weather conditions, only five MMT fields (all but the field with lowest Right Ascencion) produced usable data. This resulted in a total sky coverage of 3.92~$\rm deg^2$.\\

\begin{figure}[h]
\begin{center}
\epsfig{figure=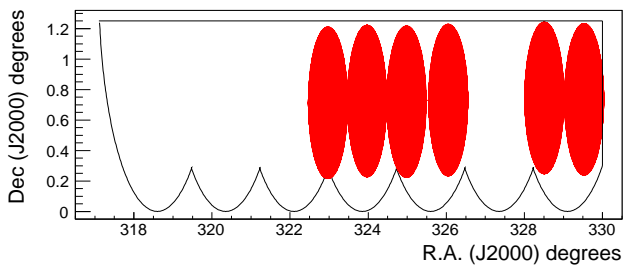,width = \columnwidth} 
\caption[]{ Footprint of the BOSS half-plates (black contours) and of the MMT fields (filled red disks) dedicated to the present study. BOSS plates were attributed the numbers 5141 to 5147 from left to right. MMT fields were labeled 0 to 5 with increasing Right Ascension.} 
\label{fig:footprint}
\end{center}
\end{figure}

Figure~\ref{fig:footprint} illustrates the locations of the BOSS and MMT fields dedicated to the present study.

\subsection{Impact of source morphology}\label{sec:morphology}

\begin{figure}[h]
\begin{center}
\epsfig{figure=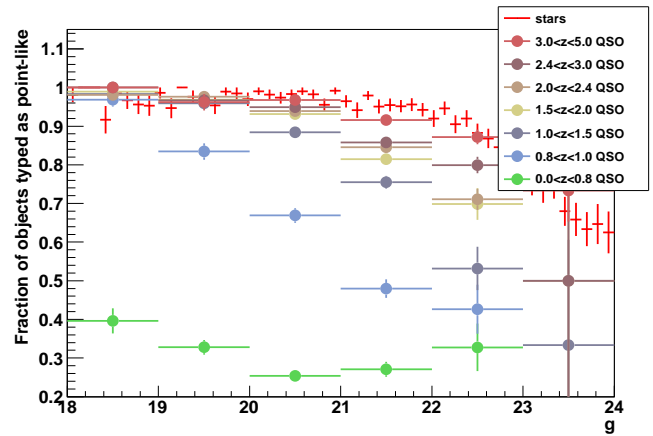,width = \columnwidth} 
\caption[]{Fraction of known quasars identified as point-like in co-added images, as a function of magnitude, for several bins in redshift (rising redshift from bottom to top). The asymptotic red histogram is the fraction of stars that are identified as point-like. } 
\label{fig:typ6}
\end{center}
\end{figure}

While most quasars, whatever their redshift,  appear point-like on single-epoch SDSS images, this is no longer the case on co-added frames, which result from the superposition, at the image level, of about 56 Stripe 82 scans, thus reaching a depth about 2 magnitudes fainter than individual scans. In the BOSS sample, we thus rejected a large fraction of low-redshift quasars by only considering point-like objects.  Fig.~\ref{fig:typ6} shows the fraction of quasars classified as point-like on the co-added images, for several bins in redshift. 

Fig.~\ref{fig:typ6} also displays the fraction of point-like objects on a single-epoch image of excellent seeing (most of these objects are indeed  stars) that are classified as stars on the co-added frame. While all stars with magnitudes $g<22$ appear point-like in the deep frame, up to 16\% of them appear extended at $g=23$, possibly due to small misalignments in the co-addition procedure that affect the object morphology at faint magnitudes,  or to noise that starts to contribute significantly in the wings of the PSF, thus smearing it out. High redshift ($z>3$) quasars follow the same trend as stars, therefore suffering from the same technical drawbacks. In contrast, as the redshift decreases, more and more quasars appear extended on the co-added frame, pointing towards a physical effect that cannot be explained with the previous arguments. Even the curves for $2.0<z<3.0$ quasars are significantly lower than that obtained for stars, our control sample of point-like objects. There is a clear step at a redshift $z\sim0.8$, where 60 to 80\% of the quasars, even bright ($g<20$) ones, appear extended. 

These results are consistent with the host galaxy becoming detectable in the co-added images, making the quasar appear extended. Fig.~\ref{fig:typ6} thus indicates a statistical detection of the host galaxy of quasars, even at redshifts as large as $z\sim2$. The brightest ($g<19$) quasars, however,  still sufficiently outshine  their host galaxy to remain point-like, except in the lowest redshift bin where the host galaxy is resolved.

\section{Completeness corrections}\label{sec:eff}

To estimate quasar densities and the luminosity function from raw quasar counts, the data have to be corrected for instrument-related losses and for biases introduced by the cuts applied to select the targets.  The completeness-corrected number of quasars is determined as
\begin{equation}
N_{QSO}  =  \sum_{N_{\rm observed}} \frac{1}{f_{\rm comp}}
\end{equation}
where ${f_{\rm comp}}$ is the fraction of recovered quasars based on the product of various selection effects.
The analysis-related completeness corrections are computed from control samples of known quasars, while the instrument-related ones are computed from the data. 

\begin{figure*}[htbp]
\begin{center}
\epsfig{figure=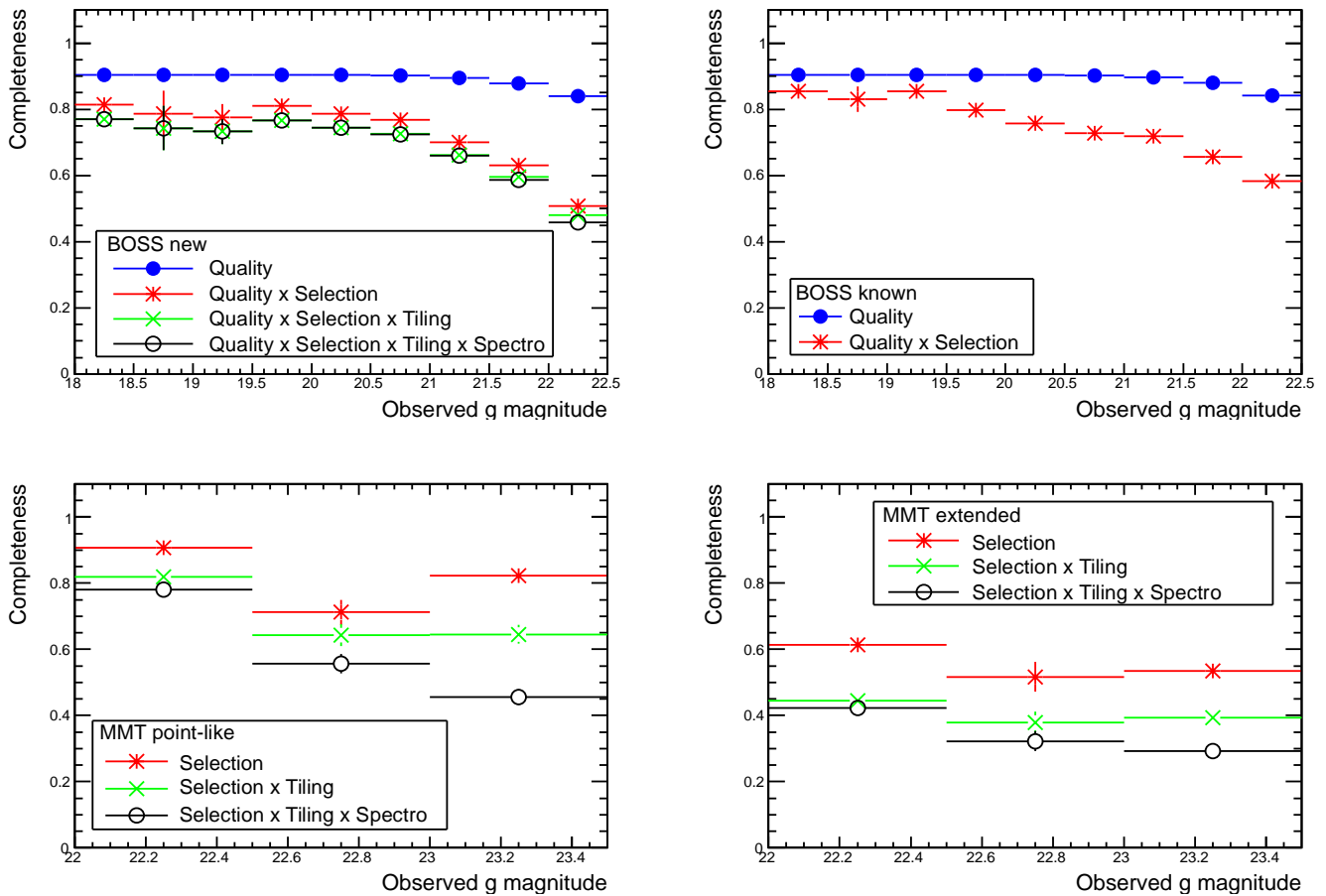,width = \textwidth} 
\caption[]{Magnitude dependence of the contributions to the total completeness correction for the different samples. Upper left: the new quasars identified in the BOSS sample. Upper right: the known quasars that were  intentionally removed  from the target lists. Lower left (resp. right): the quasars identified in the MMT sample and that appeared point-like (resp. extended) in the coadded SDSS frame. These corrections are computed from all quasars contained in either sample.} 
\label{fig:Cgeff}
\end{center}
\end{figure*}

The different contributions to the completeness corrections are detailed in the following sections, and a summary of the magnitude dependence of the corrections for BOSS and the MMT is illustrated in Fig.~\ref{fig:Cgeff}. 

\subsection{Control sample of known quasars} \label{sec:knownsample}

\begin{figure}[h]
\begin{center}
\epsfig{figure=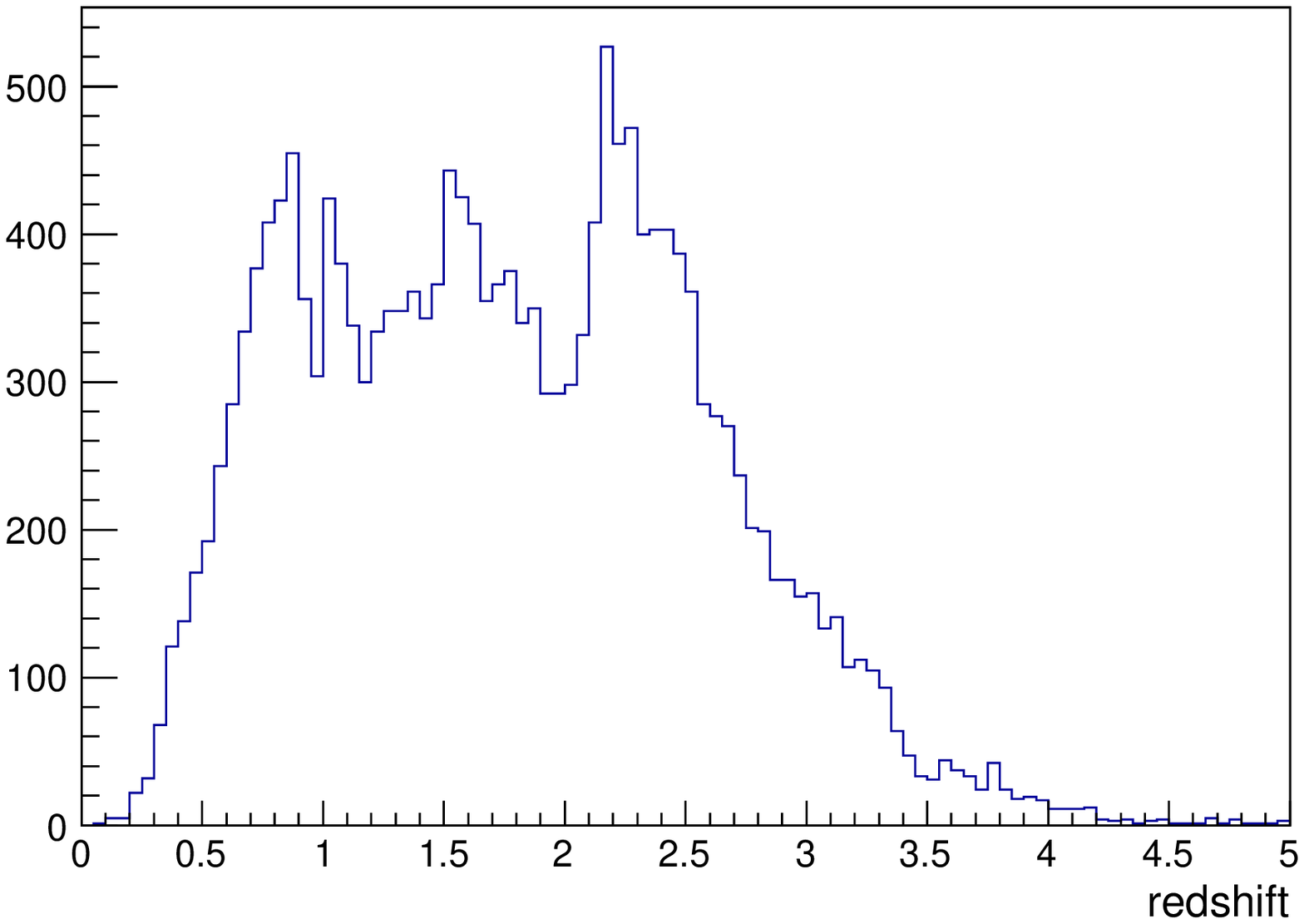,width = \columnwidth} 
\epsfig{figure=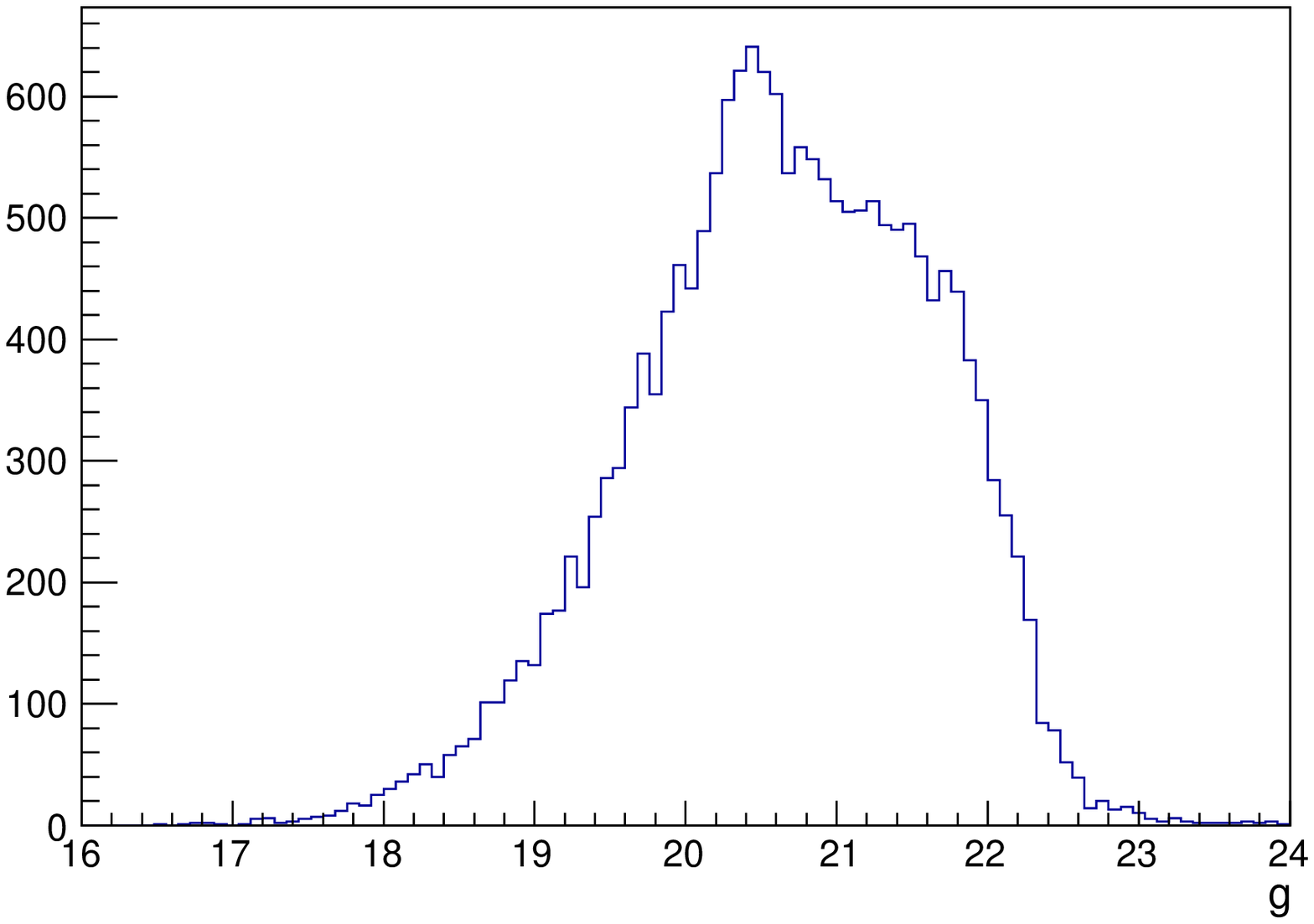,width = \columnwidth} 
\caption[]{Distributions in redshift (top plot) and extinction-corrected magnitude (bottom plot) for the sample of quasars from  previous quasar surveys covering Stripe 82. } 
\label{fig:z_distrib_known}
\end{center}
\end{figure}

The analysis-related completeness corrections are determined using a list of 19215 
spectroscopically confirmed quasars in Stripe 82 obtained from the 2dF quasar catalog 
\citep{bib:croom04}, the 2dF-SDSS LRG and Quasar Survey 2SLAQ ~\citep{bib:croom09}, the SDSS-DR7 spectroscopic database~\citep{bib:abazajian09}, the SDSS-DR7 quasar catalog~\citep{bib:schneider10} and  BOSS observations up to July 2011~\citep{bib:ahn12,bib:paris}. About 48\% of the quasars (over 90\% of the $z>2.2$ quasars and 25\% of the $z<2.2$ quasars) come from BOSS, 40\% from SDSS-DR7 and 10\% from 2SLAQ. Note that BOSS re-observed all the $z>2.2$ quasars identified in previous surveys that fell in its footprint.
As shown in Fig~\ref{fig:z_distrib_known}, these quasars have redshifts 
in the range $  0 \leq z  \leq 5 $ and $g$ magnitudes  in the  range
$17 \leq g  \leq 23$ (Galactic-extinction corrected) with 212 quasars having a magnitude fainter than  $g=22.5$. The irregular shape of the distributions results from the use of several surveys with different redshift goals and selection algorithms. At $z>2$, most of the quasars come from color-selection with the standard BOSS survey~\citep{bib:yeche10, bib:kirkpatrick11,  bib:Bovy11, bib:ross12}. 

The incompleteness of the control sample is not an issue for this analysis since it is used to compute ratios and not absolute numbers of quasars. The only hypothesis is that it covers in an unbiased way the full range of quasar parameters (such as colors, magnitudes,  morphologies, or time sampling  of the lightcurves) that the various selections depend upon.

For each of these known quasars, lightcurves  in all 5 SDSS filters were built as described in Sec.~\ref{sec:target}. We used all matching single-epoch data that passed quality criteria on the photometry described in appendix A.1 of \citet{bib:ross12}. Requiring at least one valid epoch on the lightcurve removed approximately 2\% of the objects, reducing the sample to 18910 quasars.
 The variability parameters  of Eqs \ref{eq:var_chi} and \ref{eq:var_Agamma} were computed for these quasars and used to determine completeness corrections. 

\subsection{Analysis-related completeness corrections}

The analysis-related corrections arise from two contributions that affect the BOSS and the MMT samples differently: a quality factor $\epsilon_{\rm qual}$ and a selection factor $\epsilon_{\rm sel}$. These corrections depend on the source morphology (point-like or extended), the redshift $z$ and the magnitude $g$ uncorrected for extinction (since we are here sensitive to the observed and not to the intrinsic quasar magnitude). For the sake of clarity, the magnitude corrected for Galactic extinction will henceforth be denoted $g_{\rm dered}$.

\subsubsection*{Quality completeness $\epsilon_{\rm qual}(g)$: } 
This contribution is related to the quality cuts described in the Appendix~A of~\citet{bib:ross12} that were applied to obtain the initial list of the sources from which the selection was made. It only affects BOSS data as no quality cut was applied to build the list of sources for the MMT sample (see sections \ref{sec:TS_BOSS} and \ref{sec:TS_MMT}). The mean quality factor over the BOSS data of the present study is $0.89$, with a plateau at 0.90 for all bright sources with $g < 22$, dropping to 0.70 at $g= 23$ (see blue curve in upper two plots of figure~\ref{fig:Cgeff}).

\subsubsection*{Target selection completeness\\ $\epsilon_{\rm sel} (g,\,z,\,{\rm source\,morphology})$:} 
This contribution is related to the variability-based and color-based selection algorithms, as described in Sec.~\ref{sec:target}. It depends on the telescope that the target list is designed for, because of the different  thresholds that were set on the selection variables. 

\begin{figure}[h]
\begin{center}
\epsfig{figure=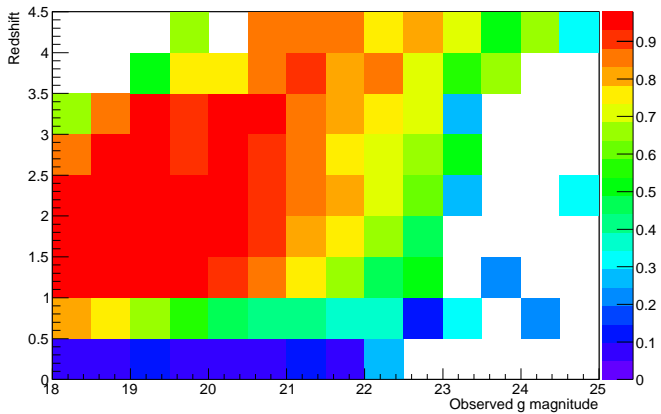,width = \columnwidth} 
\caption[]{Selection completeness  $\epsilon_{\rm sel\,BOSS} (g,\,z)$ for the quasars in the BOSS program. The significant drop at low redshift is due to the fact that many quasars with $z<1$ appear as extended sources in coadded frames and were thus not considered in the initial source list. We also observe the expected efficiency drop at faint magnitudes. Bins are left unfilled (white) are bereft of quasars.} 
\label{fig:effBOSS}
\end{center}
\end{figure}

BOSS targets were selected from an initial list limited to point-sources. To take into account the incompleteness due to the exclusion of extended sources, $\epsilon_{\rm sel\,BOSS} $ was computed from the ratio, in the control sample, of the point-like quasars that passed the selection cuts to the total number of quasars. This yielded the correction table shown in Fig.~\ref{fig:effBOSS} as a function of magnitude $g$ and redshift $z$. The mean completeness correction for the selection step, over the BOSS sample, is 0.78. 

For the MMT data,  the completeness correction for the selection step was computed separately for point-like sources and for extended ones. For each source morphology, $\epsilon_{\rm sel\,MMT} (g,\,z)$ was determined from the control sample as the fraction of quasars of a given morphology that passed the relevant selection criteria  (cf. Sec.~\ref{sec:TS_MMT}). 
The mean selection completeness correction over the MMT sample of quasars is $0.76$ ($0.82$ for the point-like sources and $0.57$ for the extended ones).


\subsection{Instrument-related completeness corrections}\label{sec:eff_spectro}

These corrections come from two contributions: a tiling factor $\epsilon_{\rm tiling}$ quantifying whether a target could indeed have been observed, and a spectrograph factor $\epsilon_{\rm spectro}$ quantifying the probability of obtaining a secure identification and redshift for a given target. These depend on the source morphology (point-like or extended) through the target priority, and on the observed magnitude $g$. 

\subsubsection*{Tiling completeness $\epsilon_{\rm tiling}({\rm priority})$:}
This contribution is related to the tiling procedure~\citep{bib:dawson12}, taking into account the density of allocated fibers and the target priorities in case of fiber collisions. This correction just represents the fraction of targets that were assigned fibers. It entirely depends on the instrumental configuration and thus must be computed specifically for BOSS and the MMT. As explained in Sec.~\ref{sec:target}, the target priority varies with magnitude for BOSS targets, and with both magnitude and source morphology (point-like or extended) for MMT targets. In the case of BOSS, however, the fraction of spectra that can be correctly identified drops significantly beyond $g\sim 22.5$ and only the first priority bin was used for the present analysis (cf. Sec.~\ref{sec:corcounts}).
The mean tiling completeness correction is $0.95$ (respectively 0.84) for the quasars recovered from the BOSS (resp. the MMT) data and selected for our measurement of the Quasar Luminosity Function. Corrections per priority bin, numbers of observed targets and of selected quasars are given in table~\ref{table:tableetil}.

\begin{table}[h]
  \begin{center}
  \begin{tabular}{cccccc}
  \hline
Project &   \multirow{2}{*}{BOSS}   & MMT   & MMT   & MMT   & MMT   \\
 (Priority) &        & A     & B     & C     & D     \\
  \hline
  $\epsilon_{\rm tiling}$ & 0.95   & 0.90  & 0.73  & 0.53  & 0.50  \\
  $N_{\rm tiled}$ & 3120 & 569& 461& 88& 78\\
  $N_{\rm QSO\, sel.}$& 860& 137& 48& 8& 0\\
  \hline
  \end{tabular}
  \end{center}
\caption{Tiling completeness correction for BOSS (unique priority level) and the MMT (priorities A through D). Row  $N_{\rm tiled}$ indicates the number of observed targets in each category, row $N_{\rm QSO\, sel.}$ indicates the number of recovered quasars selected for the determination of the quasar luminosity function  described in Sec.~\ref{sec:corcounts}.}
\label{table:tableetil}
\end{table}

\subsubsection*{Spectrograph completeness $\epsilon_{\rm spectro}({g})$:} 

\begin{figure}[h]
\begin{center}
\epsfig{figure=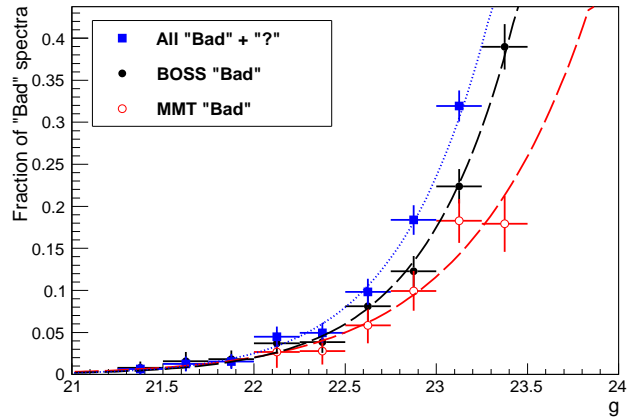,width = \columnwidth} 
\caption[]{Fraction of inconclusive spectra (declared so after visual inspection) as a function of observed $g$ magnitude. The curves are fits to the data. In our analysis, we use the upper  blue curve to correct for the total fraction of inconclusive spectra.} 
\label{fig:ratespectro}
\end{center}
\end{figure}

Some spectra did not produce a reliable identification of the source, either because the extraction procedure had failed (yielding flat and useless spectra hereafter labelled ``Bad'') or because the spectrum had too low a signal-to-noise ratio for adequate identification (hereafter ``?''). Details on the classification of the spectra can be found in~\citet{bib:paris}. 
Fig.~\ref{fig:ratespectro} illustrates the rate of the ``bad'' spectra as a function of magnitude $g$ for the BOSS (filled black dots) and the MMT (open red circles) spectra. The overlaid dashed curves are fits to the data. The fraction of ``bad'' spectra reaches 15\% at $g=23.15$ for the MMT and $g=22.9$ for BOSS. 
The upper blue dot-dashed curve illustrates the total loss $1- \epsilon_{\rm spectro}(g)$, similar for both instruments, when considering both the ``bad'' and the ``?'' spectra. This total fraction of inconclusive spectra is negligible  for $g<22$, it reaches 15\% at $g=22.8$, and 30\% at $g=23.1$.

\section{Results} \label{sec:results}

BOSS data were taken on the dedicated Sloan Foundation 2.5~m telescope~\citep{bib:gunn06} using the BOSS spectrograph~\citep{bib:smee12}. 
The BOSS spectra were reduced with the standard BOSS pipeline~(\cite{bib:bolton12, bib:dawson12}, see also \cite{bib:bolton}) which also provides an automated determination of the target classification and redshift. A manual inspection was performed on all the spectra of this program, as for all quasar targets of the main BOSS survey~\citep{bib:paris}. The MMT spectra were reduced with a customized pipeline based heavily on the E-SPECROAD package\footnote{http://iparrizar.mnstate.edu/~juan/research/ESPECROAD/}. All the spectra were checked visually to produce final identifications and redshifts. The spectra were classified as ``QSO" for secure quasars with reliable redshift, ``QSO?" for secure quasars  but uncertain redshift, ``Star", ``Galaxy" or ``Inconclusive". The latter case encompasses the ``Bad'' and the ``?''  spectra of the previous section. In our analysis, we use all spectra that were identified as ``QSO" or ``QSO?", where we call ``quasar'' an object with a luminosity $M_i[z=2] < -20.5$ and either displaying at least one emission line with FWHM greater than 500~$\rm km/s$ or, if not, having interesting/complex absorption features.

\subsection{Raw counts for the BOSS and MMT samples}

We identified $1179$ new quasars with BOSS (hereafter referred to as ``New BOSS" or simply ``BOSS") and $262$ ones with the MMT.  To these, we add the 436 previously identified quasars in our area of interest that were explicitly removed from the target lists and hereafter referred to as ``Known". Our survey data cover 14.5~${\rm deg}^2$, among which 3.92~${\rm deg}^2$ with both BOSS and MMT.

\begin{figure}[h]
\begin{center}
\epsfig{figure=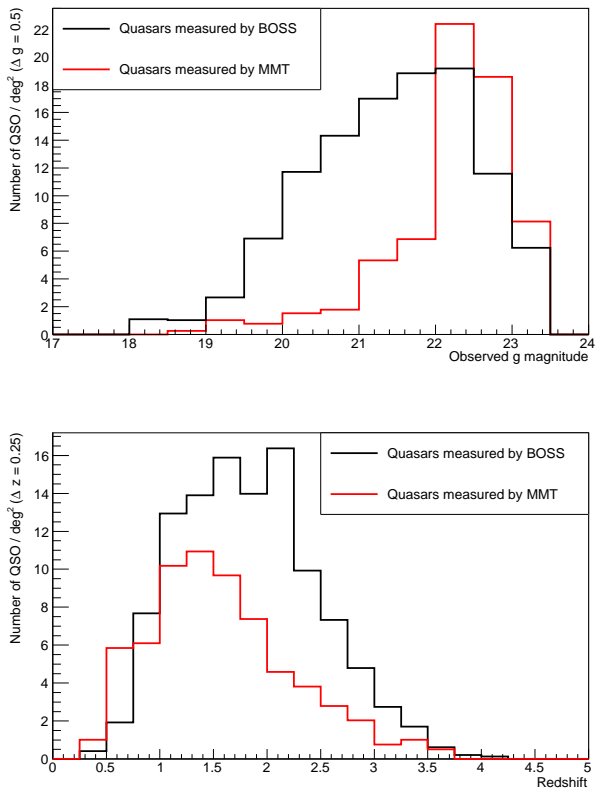,width = \columnwidth} 
\caption[]{Quasar number counts per $\rm deg^2$ as a function of observed $g$ magnitude (top) and redshift (bottom). The black (resp. red) line is for quasars measured by BOSS (resp. MMT). The BOSS histograms include the already identified quasars in the area that were removed from the target list.} 
\label{fig:cnraw}
\end{center}
\end{figure}

Fig.~\ref{fig:cnraw} shows the distributions of the redshift and observed magnitude in the $g$ band for our sample of confirmed quasars. While the redshift distributions of the BOSS and MMT samples are similar in shape, the MMT sample reaches more than half a magnitude deeper than the BOSS sample. The typical quasar magnitude of each of the two samples is $\langle g \rangle_{\rm MMT} = 22.2$ and $\langle g \rangle_{\rm BOSS} = 21.5$. The sample of previously known quasars that were not included in the target lists have a mean magnitude $\langle g \rangle_{\rm Known} = 21.0$. 

As explained in Sec.~\ref{sec:TS_MMT}, the MMT targets at $g<22$ only consisted of the selected objects that were not included in the BOSS sample, i.e., either those did not pass the quality criteria or that appeared as extended in the source catalog. This led to 69 quasars with $g<22$ in the MMT sample. For $g>22$, the MMT sample included 116 quasars in common with BOSS, plus 77 additional ones.

Table~\ref{tab:rawcounts} summarizes the raw quasar counts in the BOSS and MMT samples.
\begin{table}[htb]
\begin{center}
\begin{tabular}{l|cccc|c}
\hline
&\multicolumn{4}{c|}{Observed $g$ magnitude}&\\
Sample&$<22$&$22-22.5$&$22.5-23$&$>23$&Total\\
\hline
BOSS & 692& 231& 165& 91& 1179\\
MMT & 69& 88& 73& 32& 262\\
Known & 381& 51& 4& 0& 436\\
  \hline
  \end{tabular}
  \end{center}
\caption{Raw number counts for the different samples in several $g$ magnitude bins.}
\label{tab:rawcounts}
\end{table}

\subsection{Identification cross-checks}\label{sec:idXcheck}

We have intentionally observed some targets with both instruments in the $g>22$ magnitude regime where we expect the BOSS identification to become less secure. Table~\ref{tab:truthtable} summarizes the cross-identification of the 344 common targets, among which 116 quasars identified as such from both instruments. There are very few changes in identification. The most notable feature is that, as expected in this faint magnitude regime and compared with BOSS, MMT observations allow more quasars to be identified and given a secure redshift: 19 ``QSO?" BOSS targets were confirmed as ``QSO" with MMT, as well as 9 targets classified as ``Bad"  in BOSS. It is noteworthy that even at these faint magnitudes, there were almost no false quasar detections in BOSS: only 1 ``QSO?" BOSS target was identified as a star from the MMT spectrum; all others were confirmed as quasars.  

\begin{table}[htb]
\begin{center}
\begin{tabular}{l|ccccc}
MMT$\setminus$BOSS&
 QSO & QSO?  & Star  & Galaxy & Unknown\\
  \hline
 QSO& 94&19&3&-&9\\
 QSO?&2&1&3&-&2 \\
  Star&-&1&115&1&37 \\
 Galaxy&-&-&6&6&-\\
 Unknown&2&3&19&-&21\\
\hline
\end{tabular}
\caption[]{  Cross-identification between BOSS and MMT spectra for the 344 targets with $g>22$ that were observed by both telescopes. }
\label{tab:truthtable}
\end{center}
\end{table}

At brighter magnitudes ($22<g<22.5$), table~\ref{tab:truthtableg225} suggests excellent consistency in the identification by either telescope. 
Out of a total of 66 targets for which a secure identification is available from the MMT spectra,  59 were correctly identified by BOSS, 5 had too low S/N, and 2 were misidentified. There are no false quasar detections.  
\begin{table}[htb]
\begin{center}
\begin{tabular}{l|ccccc}
MMT$\setminus$BOSS& QSO & QSO?  & Star  & Galaxy & Unknown\\
  \hline
 QSO& 38&5&2&-&4\\
QSO?&-&-&-&-&1 \\
Star&-&-&15&-&- \\
Galaxy&-&-&-&1&-\\
Unknown&-&-&2&-&-\\
\hline
\end{tabular}
\caption[]{  Cross-identification between BOSS and MMT spectra for the 68  targets with $22<g<22.5$ that were observed by both telescopes. }
\label{tab:truthtableg225}
\end{center}
\end{table}


From the set of targets with an identification in both samples (i.e., not considering those declared ``unknown" in either case), and assuming the true identification to be the one from the MMT spectrum, we can estimate the identification reliability with BOSS to be of order 237/251 = $94\pm2\%$ at $g>22$ and of order 59/61=$97\pm 2\%$ over the magnitude range $22<g<22.5$.

\subsection{Corrected counts}\label{sec:corcounts}
To study quasar number counts, we use the quasars identified in the BOSS sample to $g<22.5$ and in the MMT sample to  $g<23.0$. The depth of the observations is insufficient to use identified quasars with $g>23.0$. The spectra are too low signal-to-noise and the spectrograph incompleteness corrections too large to yield reliable corrected counts. All fields suffer from relatively high Galactic extinction: $\langle A_g \rangle = 0.27$ (rms of 0.10) ranging from 0.15 to  0.75 in the BOSS fields and $\langle A_g \rangle= 0.25$  (rms of 0.04), ranging from 0.15 to  0.35, in the MMT fields. The previous limitations on observed magnitude therefore result in the following effective bounds on the extinction-corrected magnitude, equal to $g_{\rm max} - A_{g,\, \rm min}$: $g_{\rm dered}<22.35$ for BOSS and $g_{\rm dered} <  22.85$ for the MMT.

To compute completeness-corrected quasar counts, we defined three mutually exclusive samples,
based on observed $g$ magnitude and sky coverage:
\begin{itemize}
\item $g<22.0\;$: use of the BOSS sample over the entire $ 14.5\ \rm deg^{2}$ area (691 BOSS and 346 known, for a total of 1037 quasars)
\item $22<g<22.5\;$: use of the MMT sample over its $3.9\ \rm deg^{2}$ coverage (88 quasars)
      and of the BOSS sample for the remaining $10.6\ \rm deg^{2}$ area (169 quasars), completed by 47 known quasars
\item $22.5<g<23.0\;$ use of the MMT sample over its $3.9\ \rm deg^{2}$ coverage (105 MMT and 3 known quasars)
\end{itemize}
This division ensures, for each magnitude range, the use of the data with best redshift reliability (cf. Sec.~\ref{sec:idXcheck}) and maximal statistical significance. This also prevents us from double-counting quasars that were observed both with BOSS and the MMT. The data that did not explicitly enter the computation of the corrected counts or quasar luminosity function were used for cross-checks (cf. Secs.~\ref{sec:idXcheck} and \ref{sec: countXcheck}). 

Data from the BOSS sample were corrected for selection ($\epsilon_{\rm sel}$) and quality ($\epsilon_{\rm qual}$) incompleteness. In addition,  tiling ($\epsilon_{\rm tiling}$) and spectrograph ($\epsilon_{\rm spectro}$) completeness corrections were applied to the quasars that were identified from this deep program. The previously identified quasars in the area were only corrected by $\epsilon_{\rm sel}$ and $\epsilon_{\rm qual}$ since they were removed from the list prior to the tiling procedure.

Data from the MMT sample were corrected for spectrograph ($\epsilon_{\rm spectro}$) incompleteness, and for the relevant $\epsilon_{\rm tiling}$ and $\epsilon_{\rm sel}$ that, for the MMT sample, depended on the target morphology.

The completeness-corrected number of quasars is thus computed from the following equation:
\begin{eqnarray}
N_{QSO} & = & \sum_{N_{\rm BOSS}} \frac{1}{\epsilon_{\rm sel}\; \epsilon_{\rm qual}\;\epsilon_{\rm tiling}\;\epsilon_{\rm spectro}} \nonumber \\
                &+&\sum_{N_{\rm Known}}\frac{1}{\epsilon_{\rm  sel}\; \epsilon_{\rm  qual}}\nonumber \\
       &+& \sum_{N_{\rm MMT}^{\rm point-like}}\frac{1}{\epsilon_{\rm sel}^{\rm point-like}\; \epsilon_{\rm tiling}\;\epsilon_{\rm spectro}}
       \nonumber \\
              &+& \sum_{N_{\rm MMT}^{\rm extended}}\frac{1}{\epsilon_{\rm sel}^{\rm extended}\; \epsilon_{\rm tiling}\;\epsilon_{\rm spectro}}
\end{eqnarray}
where  the completeness corrections are 2D functions of the quasar magnitude $g$ and redshift $z$, and where the BOSS and MMT quasars entering the summations are those selected according to the 3 mutually exclusive samples defined previously.

\begin{figure}[h]
\begin{center}
\epsfig{figure=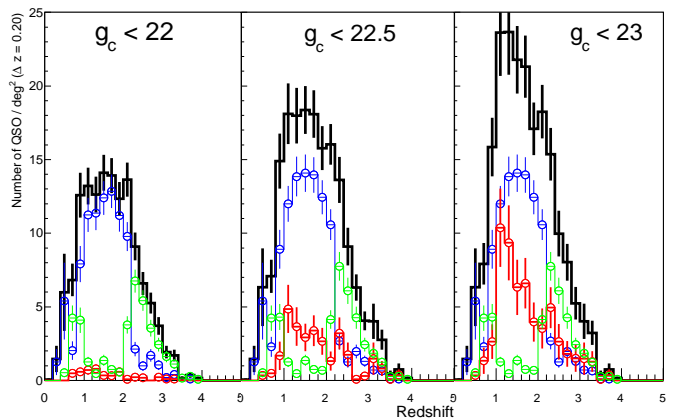,width = \columnwidth} 
\caption[]{Quasar number counts per deg$^2$ as a function of redshift ($\Delta z=0.2)$. Blue is for the new sample of quasars identified with our BOSS deep program,
green for the previously known BOSS sample,
red is for the MMT sample and black for the total. The blue, green and red curves use mutually exclusive samples and correspond to the zones defined in Sec.~\ref{sec:corcounts}.
} 
\label{fig:qsopred}
\end{center}
\end{figure}

Figure~\ref{fig:qsopred} illustrates the completeness-corrected quasar number counts as a function of redshift
for 3 magnitude limits: $g_{\rm dered}<22$ as in the BOSS survey~\citep{bib:schlegel09}, $g_{\rm dered}<22.5$ as required for the eBOSS project\footnote{http://lamwws.oamp.fr/cosmowiki/Project\_eBoss}, and $g_{\rm dered}<23$ as required for BigBOSS\footnote{http://bigboss.lbl.gov}. As explained above, however, the last set of histograms are only complete to $g_{\rm dered}<22.70$ and thus represent lower limits for $g_{\rm dered}<23$. On average over the quasars of this project, total completeness corrections are at the level of 80\% to $g<20$, and drop smoothly to 50\% at $g\sim 22.5$. 
Table~\ref{table:predzone} summarizes the total quasar number counts for
various ranges in corrected $g$ magnitude and redshift. 

\begin{table}[h]
  \begin{center}
  \begin{tabular}{c|ccccc}
  \hline

Redshift&\multicolumn{5}{c}{Extinction-corrected magnitude $g_{\rm dered}$}\\
$z$ & $< 22$ & $< 22.25$ & $< 22.5$ & $< 22.75$ & $< 23$    \\
 \hline


$<1$&          27(3)& 29(3)&  30(4)&  30(4)&  32(4)\\
$[1-2.15]$ &75(3)& 87(3)&  99(4)&  113(5)& 119(6)\\
$>2.15$&    34(2)& 37(2)&  47(3)&  49(3)& 53(3)\\
all $z$&             136(5
)&153(5)&175(6)& 191(7)& 205(8)\\
  \hline
  \end{tabular}
  \end{center}
\caption{Number counts (quasars per deg$^2$) for several ranges in extinction-corrected magnitude $g_{\rm dered}$ and redshift $z$. The statistical uncertainty on the last significant digit is indicated in parentheses. Number counts are complete to $g_{\rm dered}<22.5$, and only indicate lower limits at fainter magnitudes.}
\label{table:predzone}
\end{table}

\subsection{Counting cross-checks}\label{sec: countXcheck}

Because we had overlapping data in terms of magnitude range or sky coverage from two different programs, several counting cross-checks can be performed. Every single quasar observed was therefore used to compute the luminosity function or for cross-checks, and some for both.

\subsubsection*{Density comparison in $22<g<22.5$}

The magnitude range $22<g<22.5$ was accessible to both the MMT and the BOSS samples. Because of the different constraints on source morphology and quality, however, and since the MMT zone only covers a portion of the BOSS zone, the quasars observed in the two cases were not the same. The completeness-corrected quasar number counts derived from either sample over $22<g<22.5$ are given in table~\ref{tab:NbCountXCheck}. The last column indicates the number of identified quasars used to make the measurement. The quasar densities derived from BOSS or the MMT over this magnitude range are in excellent agreement.

\begin{table}[htb]
\begin{center}
\begin{tabular}{lcc}
\hline
\multirow{2}{*}{Survey} & QSO density& Raw number  \\ 
&  (deg$^{-2}$) & of quasars\\
\hline
BOSS over MMT zone & $39.0\;\pm\;5$ & 72 \\
BOSS complete sample &  $40.2\;\pm\;3$  & 276 \\
MMT & $39.1\;\pm\;4$ & 100\\
\hline
\end{tabular}
\caption[]{  Efficiency-corrected number counts for the BOSS and MMT samples (including, in both cases,  the BOSS standard QSOs that were selected but not re-observed) over the magnitude range $22<g<22.5$ accessible to both samples. }
\label{tab:NbCountXCheck}
\end{center}
\end{table}

\subsubsection*{Cross-check of $\epsilon_{\rm qual}$ using MMT point-like targets at $g<22$}

The sample of quasars selected for the MMT observations from point-like sources in the magnitude range $g<22$ should allow recovery of the quasars that were not selected in the BOSS target list because of the constraints on source quality. They can therefore be used to verify the estimated $\epsilon_{\rm qual}$. This concerned $N= 22$ MMT quasars, with a mean tiling correction (corresponding to priority level A) $\epsilon_{\rm tiling}= 0.903$.  Given a survey area $S= 3.9\ \rm deg^2$, the density of point-like sources observed with MMT at $g<22$ is therefore $N /\epsilon_{\rm tiling} / S = 6.2\pm1.3\ \rm deg^{-2}$.

In the same magnitude range, the BOSS sample consisted of $N = 691$ quasars, with a mean tiling correction $\epsilon_{\rm tiling}= 0.945$ and a mean source quality completeness correction $\epsilon_{\rm qual} = 0.897$. Given a survey area $S = 14.5\ \rm deg^2$, the estimated density of quasars not included in the sample because of the quality constraint is thus $N / \epsilon_{\rm tiling} / \epsilon_{\rm qual} \times (1-\epsilon_{\rm qual}) / S= 5.8\pm0.2\ \rm deg^{-2}$, in agreement within $1\sigma$ with what was estimated from the MMT sample.

\subsubsection*{Cross-check of $\epsilon_{\rm sel}$ using MMT extended targets at $g<22.5$}

Finally, the quasars selected for the MMT observations from extended sources at $g<22.5$ should allow recovery of the quasars that were not selected in BOSS because of constraints on the source morphology. This can be checked by comparing the density of extended quasars in the MMT survey to the number of quasars that were not selected in BOSS for this same reason, estimated from the observed number of point-like quasars and the morphology-part of the selection completeness correction. \begin{figure}[h]
\begin{center}
\epsfig{figure=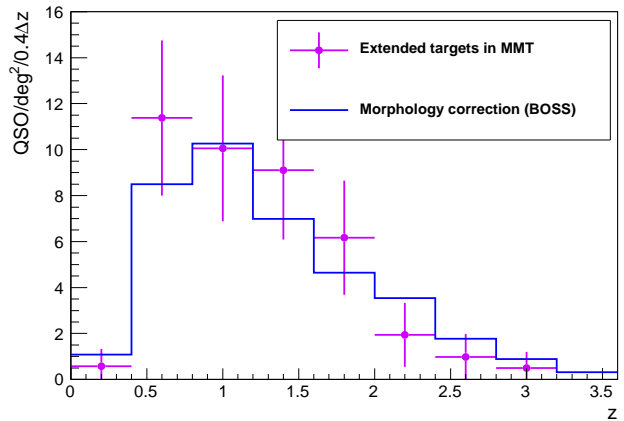,width = \columnwidth} 
\caption[]{Comparison of the number of ``extended'' quasars estimated from BOSS data and completeness corrections (blue curve) or observed in the MMT sample (purple points). 
} 
\label{fig:xcheck_typ3}
\end{center}
\end{figure}

Quantitatively, we compute the following two quantities, in the $g<22.5$ magnitude range. On the one hand, the density of extended quasars identified in MMT:
\begin{eqnarray}
&&\frac{1}{S_{\rm MMT}} \sum_{N_{\rm MMT}^{\rm extended}}\frac{1}{\epsilon_{\rm sel}^{\rm extended}\; \epsilon_{\rm tiling}\; \epsilon_{\rm spectro}}
 \nonumber
 \end{eqnarray}
and on the other hand, the density of unselected extended quasars in BOSS:
\begin{eqnarray}
&& \frac{1}{S_{\rm BOSS}} \sum_{N_{\rm BOSS}} \frac{1}{\epsilon_{\rm tiling}\; \epsilon_{\rm spectro}\; \epsilon_{\rm qual}} 
\times \left( \frac{1}{\epsilon_{\rm morphology}}-1\right) \nonumber \\
 &+& \frac{1}{S_{\rm BOSS}}  \sum_{N_{\rm Known}}\frac{1}{\epsilon_{\rm  qual}}
\times \left( \frac{1}{\epsilon_{\rm morphology}}-1\right)
 \nonumber
 \end{eqnarray}
where $S_{\rm MMT}$ and $S_{\rm BOSS} $ are respectively the areas of the MMT and the BOSS programs, and $\epsilon_{\rm morphology}$ is the fraction of the targets that are point-like (as a function of magnitude and redshift). This correction is the part of $\epsilon_{\rm sel}$ that does not include the effect of the target selection based on its color and time variability. These two densities are illustrated in Fig.~\ref{fig:xcheck_typ3}. They are clearly in agreement.

\section{Luminosity function in $g$}\label{sec:QLF}

\input{lumfct}

\section{Conclusions}
We have designed dedicated observations to measure the quasar luminosity function to $z=4$. The targets were selected with a technique relying on optical variability of the quasars, which allows both high completeness and a simple estimation of the incompleteness corrections. These include inefficiencies related to the selection technique, which we compute using a control sample of almost 20,000 known quasars, as well as instrument-related inefficiencies that we compute from the data.  The targets were shared between two instruments, the Sloan telescope (though an ancillary SDSS-III/BOSS program) and the MMT. They yielded a total of 1877 quasars, divided into  436 previously known quasars, 1179 new quasars identified with BOSS and 262 with the MMT. Cross-checks between the results from the two instruments indicate that the identification of the spectra is reliable to observed magnitudes $g\sim 23$. We have also verified that we obtained compatible number counts from BOSS and the MMT, in magnitude regimes where we had data observed with both instruments. 

These dedicated data allow us to compute reliable quasar counts and to derive the quasar  luminosity function to the limiting magnitude $g_{\rm dered} = 22.5$. Relatively high Galactic extinction in the fields prevent us from reaching fainter magnitudes. 

The quasar number counts we measure at $z<2$ are in agreement with previous estimates from \citet{bib:hopkins} or \citet{bib:croom09}. At higher redshifts, we observe of order 10\% less quasars than assumed in \citet{bib:abell09}. This trend is shown by the luminosity function that we fit to our data, corrected for incompleteness in the faintest magnitude bins using the model weighted estimator of~\citet{bib:miyaji01}. While our 
best-fit model is in good agreement with that of \citet{bib:croom09} over the common range in redshift $0.68<z<2.6$, our fainter and deeper data indicate a flatter faint luminosity slope  than what is predicted from the extrapolation of their model. This trend was already visible in the the last redshift bin  ($2.2<z<2.6$) of \citet{bib:croom09}, and has been confirmed with the present work. \citet{bib:siana08} also note a reduction in the number counts at $z\sim 3.2$ compared to LF extrapolations from low-redshift data.
At $3.5<z<4.0$, our best-fit model indicates an excess of bright quasars compared to the best-fit model of  \citet{bib:croom09}.
This result, however, is in a redshift regime where both statistical and systematic uncertainties are large. It would benefit from observations in less extincted regions of the sky, in order to reach fainter magnitudes and increased statistics.

\section*{Appendix}

Table \ref{tab:LF} provides the binned luminosity function (LF) for the data of our analysis using the model weighted estimator described in Sec.~\ref{sec:QLF}, as plotted in Fig.~\ref{fig:clum}. We give the value of $\log \Phi$ in 8 redshift intervals from $z=0.68$ to $z=4.00$, and for $\Delta M_g = 0.40$ magnitude bins from $M_g=-29$ to $M_g=-21$ to . We also give the number of quasars ($N_Q$) contributing to the LF in the bin and the error ($\Delta \log\Phi$).
\begin{table*}[h]
  \begin{center}
  \begin{tabular}{c|ccc|ccc|ccc|ccc}
  \hline
  $M_g$ & \multicolumn{3}{c|}{$0.68<z<1.06$} &  \multicolumn{3}{c|}{$1.06<z<1.44$}&  \multicolumn{3}{c|}{$1.44<z<1.82$}&  \multicolumn{3}{c}{$1.82<z<2.20$}\\
(bin center)  & $N_Q$& $\log\Phi$&$\Delta\log\Phi$& $N_Q$& $\log\Phi$&$\Delta\log\Phi$& $N_Q$& $\log\Phi$&$\Delta\log\Phi$& $N_Q$& $\log\Phi$&$\Delta\log\Phi$\\
  \hline
\input{LF14}
  \hline
$M_g$ & \multicolumn{3}{c|}{$2.20<z<2.60$} &  \multicolumn{3}{c|}{$2.60 <z<3.00$}&  \multicolumn{3}{c|}{$3.00 <z<3.50$}&  \multicolumn{3}{c}{$3.50 <z<4.00$}\\
  (bin center)   & $N_Q$& $\log\Phi$&$\Delta\log\Phi$& $N_Q$& $\log\Phi$&$\Delta\log\Phi$& $N_Q$& $\log\Phi$&$\Delta\log\Phi$& $N_Q$& $\log\Phi$&$\Delta\log\Phi$\\
  \hline
 \input{LF58}
  \hline
\end{tabular}
  \end{center}
\caption{Binned luminosity function using the model weighted estimator. The 1367 quasars used here are those with $g_{\rm dered}<22.5$ and resulting from the division into mutually exclusive samples as described in Sec.~\ref{sec:corcounts} and shown in Fig.~\ref{fig:qsopred}.  }
\label{tab:LF}
\end{table*}

\begin{acknowledgements}
The observations reported here were obtained in part at the MMT Observatory, a facility operated jointly by 
the Smithsonian Institution and the University of Arizona. \\
The other observations were obtained as part of the SDSS-III/BOSS project. 
Funding for SDSS-III has been provided by the Alfred P. Sloan Foundation, the Participating Institutions, the National Science Foundation, and the U.S. Department of Energy Office of Science. The SDSS-III web site is http://www.sdss3.org/.\\
SDSS-III is managed by the Astrophysical Research Consortium for the Participating Institutions of the SDSS-III Collaboration including the University of Arizona, the Brazilian Participation Group, Brookhaven National Laboratory, University of Cambridge, Carnegie Mellon University, University of Florida, the French Participation Group, the German Participation Group, Harvard University, the Instituto de Astrofisica de Canarias, the Michigan State/Notre Dame/JINA Participation Group, Johns Hopkins University, Lawrence Berkeley National Laboratory, Max Planck Institute for Astrophysics, Max Planck Institute for Extraterrestrial Physics, New Mexico State University, New York University, Ohio State University, Pennsylvania State University, University of Portsmouth, Princeton University, the Spanish Participation Group, University of Tokyo, University of Utah, Vanderbilt University, University of Virginia, University of Washington, and Yale University. \\
The French Participation Group to SDSS-III is supported by the Agence Nationale de la Recherche under grant ANR-08-BLAN-0222. 
N.P.-D. and Ch.Y. acknowledge support from grant ANR-11-JS04-011-01. 
A.D.M. is a research fellow of the Alexander von Humboldt Foundation of Germany. 
X.F. and I.D.M. acknowledge supports from a David and Lucile Packard Fellowship, and NSF Grants AST 08-06861 and AST 11-07682.
\end{acknowledgements}

\end{document}

%% file: lumfct.tex
We compute the quasar luminosity function (LF) from the corrected number counts derived above, and considering our completeness limit at $g_{\rm dered}<22.5$.
The distance modulus $d_M(z)$ is computed using the standard flat $\Lambda$CDM model
with the cosmological parameters of \cite{bib:larson11}: $\Omega_M=0.267$,  $\Omega_\Lambda=0.734$ and $h=0.71$.

\subsection{K-corrections}

\begin{figure}[h]
\begin{center}
\epsfig{figure=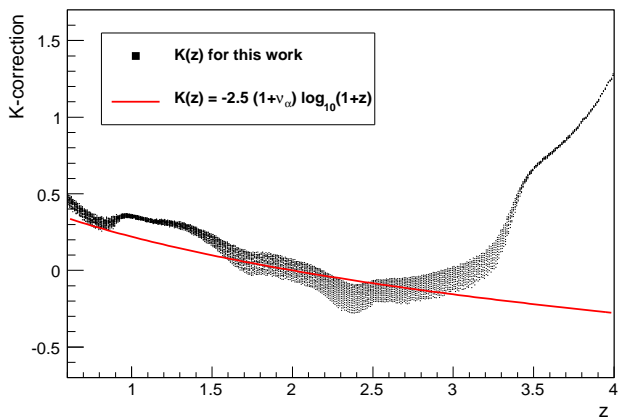,width = \columnwidth} 
\caption[]{K-corrections as a function of redshift. The spread illustrates the variation over the luminosity range of the quasars in our sample. The red curve shows for comparison the  parameterization $K(z)=-2.5 (1+\nu_\alpha) \log_{10}(1+z)$ introduced in C09, with $\nu_\alpha=-0.5$  and normalized to $z=2$.} 
\label{fig:kcorr}
\end{center}
\end{figure}

%
Selection for this survey was performed in the $g$-band, and for the
majority of the data this band provides the highest $S/N$. We define
the $K$-correction in terms of the observed $g$ magnitude and follow
\citet{bib:croom09} (hereafter C09) in applying the correction relative to $z=2$, which is
near the median redshift of our quasar sample (see also
\citet{bib:richards09}). The absolute magnitude normalized to $z=2$ is
given by:
\begin{equation}
M_g(z=2)=g_c-d_M(z)-[K(z)-K(z=2)]\;.
\end{equation}
Hereafter we will use $M_g$ as a shorthand for the redshift-corrected
$M_g(z=2)$. The $K$-correction as a function of redshift is derived
from model quasar spectra in a similar fashion to \citet{bib:richards09}.
The quasar model includes a broken power law continuum with
$\alpha_\nu=-0.5$ at $\lambda>1100\AA$ and $\alpha_\nu=-1.5$ at
$\lambda \le 1100 \AA$ \citep{bib:telfer02}. Strong quasar emission lines
are included, where the equivalent width is a function of luminosity
according to the well-known Baldwin Effect \citep{bib:baldwin77}; thus,
the $K$-correction is a function not only of redshift but also
luminosity (or equivalently, observed $g$-magnitude). The model also
includes Fe emission using the template of \citet{VW01} and Lyman-$\alpha$ forest absorption using the prescription of \citet{bib:worseck}. The
forest model is particularly relevant here, as for $z\ga2.5$ the
$g$-band $K$-correction necessarily includes a component due to forest
absorption; our $K$-correction accounts for the mean value but for
high redshift objects the uncertainty in individual $K$-corrections is
increased by line-of-sight fluctuations in the amount of forest
absorption within the $g$-band. The models used to derive the
$K$-corrections will be described in fuller detail in McGreer et al.
2012 (in prep). In general, as shown in Fig.~\ref{fig:kcorr}, the values are very similar to those used
by C09 (e.g., their Fig. 1), but are extended to $z=4$. At
$z\sim2-3$,  our use of a luminosity-dependent $K$-correction introduces 
a spread in $K$-correction values of $\sim 0.25$ mag across the luminosity range of the
quasars in our sample, as the Lyman-$\alpha$ and C IV lines are
within the $g$-band and contribute substantially to the flux within
the bandpass. 

\subsection{Luminosity function determination} \label{sec:LFmodel}
We define eight redshift bins: the first five are the same redshift intervals as in C09, with
limits $0.68,1.06,1.44,1.82,2.2,2.6$;
the last three are specific to our analysis and have the  limits $2.6,3.0,3.5,4.0$.

 For our quasar sample, we calculate the binned LF using the model-weighted estimator $\Phi$ suggested by \citet{bib:miyaji01}, which  presents the advantage of not having to assume a uniform distribution across each bin, unlike 1/V estimators. Instead, it models the unbinned LF data and uses it to correct
for the variation of the LF and for the completeness within each bin, which is here particularly critical at the faint end of the LF where the latter is incompletely sampled. This estimator gives the binned LF as
\begin{equation}
\Phi(M_{g_i},z_i)\;=\;\Phi^{model}(M_{g_i},z_i)\;\frac{N_i^{obs}}{N_i^{model}}\; ,
\label{eq:phimodel}
\end{equation}
where $M_{g_i}$ and $z_i$ are, respectively, the absolute magnitude and the redshift at the center of bin $i$,
$\Phi^{model}$ is the model LF estimated at the center of the bin,
$N_i^{model}$ is the number of quasars with $g_{\rm dered}<22.5$ estimated from the model in the bin and $N_i^{obs}$ is the observed number of quasars in the bin. 
A drawback of this estimator is that it is model-dependent, but \citet{bib:miyaji01} show that the uncertainties  due to the model dependence are practically negligible. 

We assume the LF to be appropriately described by a standard double power-law of the form~\citep{bib:boyle00}:
\begin{equation}
\Phi(M_g,z)\;=\;\frac{\Phi^*}{10^{0.4(\alpha+1)(M_g-M_g^*)}\;+\;10^{0.4(\beta+1)(M_g-M_g^*)}}
\label{eq:phi}
\end{equation}
where $\Phi$ is the quasar comoving space density.
For our $\Phi^{model}$, we follow the pure luminosity evolution model of C09, where a redshift dependence is introduced through an evolution in $M_g^*$  described by 
\begin{equation}
M_g^*(z)=M_g^*(0)-2.5(k_1z+k_2z^2)\; .
\label{eq:mg}
\end{equation}

\begin{figure*}[htbp]
\begin{center}
\epsfig{figure=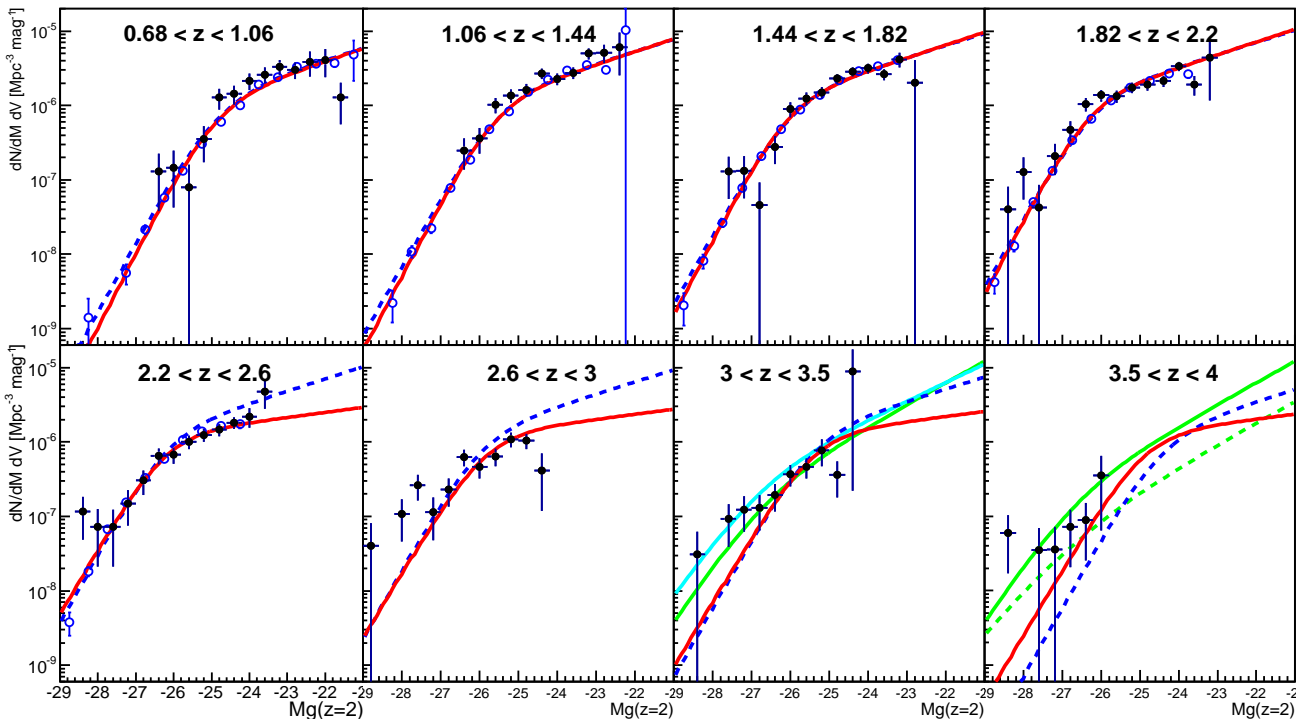,width = \textwidth} 
\caption[]{
Quasar luminosity function measurements (black circles)
compared to the C09 data (blue open circles at $0.68<z<2.6$). The blue dashed curve is the LF of C09  
and the best-fit model of this analysis is shown as the red curve. The green curves are best fits to COSMOS data~\citep{bib:masters12} at $z\sim 3.2$ (plain curve) and $z\sim 4.0$ (dashed curve). The cyan line is the best fit,  with fixed-$\alpha$, to binned LF data from SWIRE and SDSS at $z\sim 3.2$~\citep{bib:siana08}.
} 
\label{fig:clum}
\end{center}
\end{figure*}

Systematics on our LF measurements are computed by estimating the uncertainty on the completeness corrections of Sec.~\ref{sec:eff} using Monte Carlo simulations in which the selected and total number of quasars follow Poisson distributions centered on the observed means. The error bars on each measurement include both our statistical and systematic uncertainties, added in quadrature.

\subsection{Comparison to luminosity function of Croom et al. (2009) }
 In a first step, we keep all  6 parameters ($\Phi^*$, $\alpha$, $\beta$, $ M_g^*$, $k_1$ and $k_2$) of the LF fixed.
We impose, for our model $\Phi^{model}$, the values fitted by C09 for the $0.4<z<2.6$ redshift range and given in their table 2, ie. $M_g^*(0) = -22.17$, $k_1=1.46$ and $k_2=-0.328$, and we compare our data to this luminosity function.
Figure~\ref{fig:clum} illustrates our LF measurements (black circles).
The blue open circles are the LF measurements from C09
and the blue dashed curve is their best-fit $\Phi^{model}$.

The data presented in this work from dedicated BOSS and MMT surveys cover the redshift range from $z=0.68$ to $z=4$, allowing 3 additional redshift bins beyond those of C09. 
Over the $0.68<z<2.6$  redshift range common to both studies, our data and the C09 points are in good agreement.  The extrapolation of the C09 model to $z=4$ is also in reasonable agreement with our data points. Two discrepancies, however, are worth being noted. In the $3.5<z<4.0$ redshift bin, our measured LF points are all above the extrapolated curve. The low number of quasars with $z>3.5$ in our control sample, however, yields large uncertainties in the completeness corrections that result in systematic uncertainties on our LF measurements in this redshift regime. Moreover, the statistical uncertainties on the measured LF are also very large and the trend needs to be confirmed with improved statistics. 
In the $2.2<z<2.6$ bin, the C09 LF measurements are all systematically below the best-fit C09 model and this trend is corroborated by our data in the same redshift bin, confirming the hypothesis that the feature is real and not related to a redshift limitation of the sample since ours extends significantly beyond $z=2.6$. This disagreement was already noted by C09 who observed a significant improvement of  the fit when reducing the upper limit in redshift to $z<2.1$.

\subsection{Luminosity function fit}\label{sec:fitLF}
In a second step, since the data from our analysis and from C09 are independent, we included both samples to constrain the luminosity function in larger redshift and magnitude ranges than either data set alone.
 In the fitting procedure, we let all parameters of the model free and do a least-squares-fit, using the MINUIT package~\citep{bib:minuit}, to determine the best fit values and their errors. Since our LF measurements depend on the fitted $\Phi^{model}$, iterative fitting is performed to determine the luminosity function of our sample from Eq.~\ref{eq:phimodel}, until parameter convergence is reached. Up to $z\sim 2.6$, the choice of model used in this procedure only changes the last magnitude bin since it's the only one affected by the $g_{\rm dered}$ cut. Above $z\sim 2.6$, the cut on $g_{\rm dered}$ typically affects the last two magnitude bins, where the LF measurement and its error can change by up to a factor of 3.  

Fitting the model defined in Eqs.~\ref{eq:phi} and \ref{eq:mg} to the combined LF does not significantly change  the result from the best-fit model of C09. In particular, Eqs.~\ref{eq:phi} and \ref{eq:mg}  do not provide sufficient freedom to solve the discrepancies mentioned above. 

We therefore defined a new model, based on the same equations as before but allowing the redshift evolution parameters ($k_1$ and $k_2$) and the model slopes ($\alpha$ and $\beta$) to be different on either side of a pivot redshift $z_p=2.2$. The model is thus described by Eq.~\ref{eq:phi} where $\alpha$ and $\beta$ now have subscripts $l$ for $z<z_p$ and $h$ for $z>z_p$, and an evolution in $M_g^*$  characterized by 
\begin{equation}
M_g^*(z)=M_g^*(z_p)-2.5[k_1(z-z_p)+k_2(z-z_p)^2]\; ,
\label{eq:mg2}
\end{equation}
where $k_1$ and $k_2$ are again to be considered separately for low (subscript $l$) and high (subscript $h$) redshifts w.r.t. $z_p$. This more general form of our model now contains 10 parameters that are all let free.

\begin{table*}[htb]
\begin{center}
\begin{tabular}{cccccccc}
\hline
$M_g^*(z_p)$ & $\log (\Phi^*)$ & $\alpha_l$ & $\beta_l$ & $k_{1l}$ & $k_{2l}$   \\
\hline
 -26.23  & -5.84&-3.33 & -1.41 & 0.02 & -0.33\\
-26.36$\pm$ 0.06& -5.89$\pm$ 0.03&-3.50$\pm$0.05 & -1.43$\pm$0.03 &  0.03$\pm$ 0.02& -0.34$\pm$ 0.01\\
  \hline
 \end{tabular}
 \begin{tabular}{cccccc}
 \\
  \hline
  $\alpha_h$ & $\beta_h$ &$k_{1h}$ & $k_{2h}$ \\
\hline
-3.33 & -1.41 &0.02 & -0.33\\
-3.19$\pm$ 0.07& -1.17$\pm$0.05& -0.35$\pm$ 0.13& -0.02$\pm$ 0.14\\
\hline
  \end{tabular}
  \end{center}
\caption{Values of the parameters of the quasar luminosity function (Eqs.~\ref{eq:phi} and \ref{eq:mg2}). The normalization of the luminosity function ($\Phi^*$) is given in $\rm Mpc^{-3}\;mag^{-1}$. The top line gives the parameters published in C09 recalculated according to the definition of Eq.~\ref{eq:mg2} and used to initialize our fit; the bottom line is for our best-fit model on the combined data set.
}
\label{tab:lumi}
\end{table*}

The best-fit  parameters are reported in Table~\ref{tab:lumi}, and the resulting luminosity function is illustrated in Fig.~\ref{fig:clum} as the red curve. The high redshift parameters result from a fit to the data at $z>z_p = 2.2$. The faint end slope $\beta_h$ is well constrained by the $2.2<z<2.6$ data points. Given the large error bars on the data at $z>3$, however, the redshift evolution is much less constrained, as reflected by the large error bars on $k_{1h}$ and $k_{2h}$.
 The best fit has a $\chi^2$ of 324 for 162 degrees of freedom. For the sake of comparison, we also provide the parameters of C09, translated into the parameters defined in  Eq.~\ref{eq:mg2} with a  pivot redshift $z_p=2.2$ instead of $z_p=0$ in the original work of C09.
We provide, in Appendix A, the measured luminosity function values and associated uncertainties, where the measurements were corrected using the estimator described in Sec.~\ref{sec:LFmodel} and considering the best-fit model.

In the redshift bins from $z=0.68$ to $z<2.2$ where the data from C09 are the most constraining, the best-fit model shows negligible difference with that of C09. It starts to deviate at redshifts $z>2.2$, with a significantly flatter faint luminosity slope (smaller $|\beta|$) than observed at low redshift. Our data, however, only constrain the faint luminosity slope to $z\sim 2.6$, as the $2.6<z<3.0$  redshift bin has only 2 points  beyond the knee, and the following bins have none with small error bars. Deeper data should be used for a secure measurement of the high-redshift faint-luminosity slope. A similar trend, however, was observed by \citet{bib:siana08} at $z= 3.2$, where the authors noted that estimates at high redshift that were derived at low redshift and normalized using bright high-redshift QSOs resulted in a factor of about 2 overestimate at $z\sim 3$. 

Further discussions on the quasar luminosity function can be found in \citet{bib:ross12b}, where the BOSS data of this analysis are combined with the full data set from the DR9 release of the SDSS-III/BOSS survey. The quasar selection there requires models of quasar colors to determine the selection completeness, but the statistics are increased by over an order of magnitude, thus allowing better constrains on the LF model in the high redshift bins.

Using our best-fit parameters of the quasar luminosity function, we provide, in Table~\ref{tab:counts}, an estimate of the number of quasars in the redshift-magnitude plane for an hypothetical survey covering $10,000$~deg$^2$. Integrated to $g=23$ as expected for instance for the future BigBOSS survey, these counts indicate a $\sim10\%$ reduction compared to the extrapolation  of the \citet{bib:hopkins} luminosity function that was used to predict quasar number counts in the LSST science book\footnote{http://www.lsst.org/files/docs/sciencebook} \citep{bib:abell09}.
\begin{table*}[htb]
\begin{center}
\begin{tabular}{cccccccc}
\hline
$g\setminus z$ & 0.5 & 1.5 & 2.5 & 3.5 & 4.5 & 5.5 & Total \\
  \hline

15.75 & 76 & 15 & 0 & 0 & 0 & 0 & 92 \\ 
16.25 & 174 & 55 & 11 & 0 & 0 & 0 & 239 \\ 
16.75 & 402 & 172 & 61 & 0 & 0 & 0 & 635 \\ 
17.25 & 939 & 535 & 180 & 6 & 0 & 0 & 1661 \\ 
17.75 & 2163 & 1630 & 508 & 21 & 1 & 0 & 4323 \\ 
18.25 & 4740 & 4720 & 1409 & 57 & 2 & 0 & 10928 \\ 
18.75 & 9456 & 12380 & 3784 & 156 & 5 & 0 & 25781 \\ 
19.25 & 16612 & 27796 & 9409 & 422 & 14 & 0 & 54255 \\ 
19.75 & 25537 & 51561 & 20579 & 1128 & 39 & 1 & 98846 \\ 
20.25 & 35185 & 80209 & 38096 & 2923 & 107 & 4 & 156523 \\ 
20.75 & 45008 & 110341 & 59939 & 7085 & 289 & 10 & 222671 \\ 
21.25 & 54980 & 141918 & 82650 & 15386 & 779 & 27 & 295740 \\ 
21.75 & 64988 & 176959 & 103733 & 28916 & 2036 & 74 & 376706 \\ 
22.25 & 74189 & 217815 & 122861 & 46636 & 5064 & 201 & 466766 \\ 
22.75 & 80370 & 266716 & 141310 & 65652 & 11408 & 545 & 566001 \\ 
23.25 & 79024 & 325945 & 160621 & 82972 & 22419 & 1436 & 672417 \\ 
23.75 & 61347 & 398006 & 182048 & 97320 & 37756 & 3632 & 780110 \\ 
24.25 & 15976 & 480676 & 206510 & 109295 & 55090 & 8401 & 875949 \\ 
24.75 & 0 & 492283 & 234874 & 120118 & 71481 & 17111 & 935866 \\ 
\hline
Total & 571169 & 2789734 & 1368583 & 578092 & 206489 & 31444 & 5545510 \\ 

\hline
 \end{tabular}
  \end{center}
\caption{Predicted number of quasars over $15.5<g<25$ and $0<z<6$ for a survey covering 10,000 deg$^2$, based on our best-fit luminosity function. Bins are centered on the indicated magnitude and redshift values. The ranges in each bin are $\Delta g=0.5$ and $\Delta z = 1$.}
\label{tab:counts}
\end{table*}


%% file: LF14.tex
-28.80 & 0 & - & - & 0 & - & - & 0 & - & - & 0 & - & - \\
-28.40 & 0 & - & - & 0 & - & - & 0 & - & - & 1 & -7.40 & 0.43 \\
-28.00 & 0 & - & - & 0 & - & - & 0 & - & - & 3 & -6.89 & 0.25 \\
-27.60 & 0 & - & - & 0 & - & - & 3 & -6.88 & 0.25 & 1 & -7.37 & 0.43 \\
-27.20 & 0 & - & - & 0 & - & - & 3 & -6.88 & 0.25 & 5 & -6.68 & 0.19 \\
-26.80 & 0 & - & - & 0 & - & - & 1 & -7.34 & 0.43 & 11 & -6.33 & 0.13 \\
-26.40 & 2 & -6.88 & 0.31 & 5 & -6.60 & 0.19 & 6 & -6.56 & 0.18 & 24 & -5.98 & 0.09 \\
-26.00 & 2 & -6.84 & 0.31 & 7 & -6.44 & 0.16 & 19 & -6.05 & 0.10 & 31 & -5.86 & 0.08 \\
-25.60 & 1 & -7.10 & 0.44 & 19 & -5.99 & 0.10 & 26 & -5.91 & 0.09 & 29 & -5.88 & 0.08 \\
-25.20 & 4 & -6.45 & 0.22 & 24 & -5.87 & 0.09 & 30 & -5.82 & 0.08 & 36 & -5.76 & 0.07 \\
-24.80 & 11 & -5.89 & 0.13 & 27 & -5.79 & 0.08 & 43 & -5.64 & 0.07 & 36 & -5.72 & 0.07 \\
-24.40 & 13 & -5.84 & 0.12 & 42 & -5.57 & 0.07 & 49 & -5.54 & 0.06 & 36 & -5.67 & 0.07 \\
-24.00 & 17 & -5.67 & 0.11 & 33 & -5.64 & 0.08 & 49 & -5.50 & 0.06 & 51 & -5.47 & 0.06 \\
-23.60 & 18 & -5.58 & 0.11 & 35 & -5.56 & 0.08 & 39 & -5.58 & 0.07 & 16 & -5.72 & 0.13 \\
-23.20 & 22 & -5.48 & 0.10 & 49 & -5.30 & 0.07 & 30 & -5.38 & 0.10 & 2 & -5.36 & 0.32 \\
-22.80 & 17 & -5.52 & 0.11 & 33 & -5.29 & 0.12 & 1 & -5.69 & 0.45 & 0 & - & - \\
-22.40 & 18 & -5.41 & 0.16 & 8 & -5.22 & 0.25 & 0 & - & - & 0 & - & - \\
-22.00 & 14 & -5.39 & 0.18 & 0 & - & - & 0 & - & - & 0 & - & - \\
-21.60 & 4 & -5.89 & 0.25 & 0 & - & - & 0 & - & - & 0 & - & - \\
-21.20 & 0 & - & - & 0 & - & - & 0 & - & - & 0 & - & - \\

%% file: LF58.tex
-28.80 & 0 & - & - & 1 & -7.39 & 0.44 & 0 & - & - & 0 & - & - \\
-28.40 & 3 & -6.94 & 0.25 & 0 & - & - & 1 & -7.51 & 0.44 & 2 & -7.22 & 0.31 \\
-28.00 & 2 & -7.14 & 0.31 & 3 & -6.96 & 0.25 & 0 & - & - & 0 & - & - \\
-27.60 & 2 & -7.14 & 0.31 & 7 & -6.58 & 0.16 & 3 & -7.04 & 0.25 & 1 & -7.46 & 0.44 \\
-27.20 & 4 & -6.83 & 0.22 & 3 & -6.94 & 0.25 & 4 & -6.91 & 0.22 & 1 & -7.45 & 0.44 \\
-26.80 & 8 & -6.52 & 0.15 & 6 & -6.64 & 0.18 & 4 & -6.89 & 0.22 & 2 & -7.14 & 0.31 \\
-26.40 & 17 & -6.18 & 0.11 & 16 & -6.20 & 0.11 & 6 & -6.71 & 0.18 & 2 & -7.05 & 0.31 \\
-26.00 & 17 & -6.17 & 0.11 & 11 & -6.34 & 0.13 & 10 & -6.43 & 0.14 & 2 & -6.45 & 0.36 \\
-25.60 & 24 & -5.99 & 0.09 & 14 & -6.19 & 0.12 & 11 & -6.34 & 0.13 & 0 & - & - \\
-25.20 & 28 & -5.90 & 0.08 & 21 & -5.96 & 0.10 & 9 & -6.11 & 0.17 & 0 & - & - \\
-24.80 & 30 & -5.83 & 0.08 & 19 & -5.98 & 0.10 & 4 & -6.44 & 0.22 & 0 & - & - \\
-24.40 & 33 & -5.74 & 0.08 & 3 & -6.39 & 0.31 & 2 & -5.05 & 0.42 & 0 & - & - \\
-24.00 & 18 & -5.66 & 0.13 & 0 & - & - & 0 & - & - & 0 & - & - \\
-23.60 & 7 & -5.32 & 0.18 & 0 & - & - & 0 & - & - & 0 & - & - \\
-23.20 & 0 & - & - & 0 & - & - & 0 & - & - & 0 & - & - \\
-22.80 & 0 & - & - & 0 & - & - & 0 & - & - & 0 & - & - \\
-22.40 & 0 & - & - & 0 & - & - & 0 & - & - & 0 & - & - \\
-22.00 & 0 & - & - & 0 & - & - & 0 & - & - & 0 & - & - \\
-21.60 & 0 & - & - & 0 & - & - & 0 & - & - & 0 & - & - \\
-21.20 & 0 & - & - & 0 & - & - & 0 & - & - & 0 & - & - \\